\documentstyle[procl,epsfig,psfig]{article}

\input{epsf}

\def\Journal#1#2#3#4{{#1} {\bf #2}, #3 (#4)}

\def\NPB{{\em Nucl. Phys.} B}
\def\PLB{{\em Phys. Lett.}  B}
\def\PRL{\em Phys. Rev. Lett.}
\def\PRD{{\em Phys. Rev.} D}
\def\ZPC{{\em Z. Phys.} C}

\newcommand{\X}{\mbox{$X$}}
\newcommand{\Y}{\mbox{$Y$}}
\newcommand{\mx}{\mbox{$M_X$}}
\newcommand{\my}{\mbox{$M_Y$}}
\newcommand{\mpr}{\mbox{$M_p$}}
\newcommand{\x}{\mbox{$x$}}
\newcommand{\W}{\mbox{$W$}}
\newcommand{\ttra}{\mbox{$t$}}
\newcommand{\modt}{\mbox{$|t|$}}
\newcommand{\qsq}{\mbox{$Q^2$}}
\newcommand{\wsq}{\mbox{$W^2$}}
\newcommand{\xpom}{\mbox{$x_{I\!\! P}$}}
\newcommand{\fdd}{\mbox{$F_2^{D}$}}

\newcommand{\fddt}{\mbox{$F_2^{D(3)}$}}
\newcommand{\fddtild}{\mbox{${\tilde{F}_2^D}$}}

\newcommand{\albar}{\mbox{$\bar \alpha$}}

\newcommand{\gevcsq}{\mbox{$\rm GeV/c^2$}}
\newcommand{\gevsq}{\mbox{${\rm GeV}^2$}}
\newcommand{\gevsqm}{\mbox{${\rm GeV}^{-2}$}}

\newcommand{\av}[1]{\mbox{$ \langle #1 \rangle $}}
\newcommand{\lsim}{\raisebox{-0.5mm}{$\stackrel{<}{\scriptstyle{\sim}}$}}
\newcommand{\gsim}{\raisebox{-0.5mm}{$\stackrel{>}{\scriptstyle{\sim}}$}}

\def\ra{\rightarrow}

\def\al{\alpha}

\def\be{\begin{equation}}
\def\ee{\end{equation}}
\def\bea{\begin{eqnarray}}
\def\eea{\end{eqnarray}}

\newcommand {\alphapom} {\mbox{$\alpha_{_{\pom}}$}}
\newcommand {\pom}  {I\hspace{-0.2em}P}

\begin{document}

\hspace*{\fill}\parbox[t]{4cm}{DESY 96-179}


\title{DIFFRACTIVE INTERACTIONS}


\author{V. DEL DUCA}
\address{Particle Physics Theory Group,\,
Dept. of Physics and Astronomy\\ University of Edinburgh,\,
Edinburgh EH9 3JZ, Scotland, UK}

\author{E. GALLO}
\address{INFN Firenze, Largo E. Fermi 2,
50125 Firenze, Italy}

\author{P. MARAGE}
\address{Universit\'e Libre de Bruxelles, Bd. du Triomphe,
B-1050 Brussels, Belgium}


\maketitle\abstracts{
The general framework of diffractive deep inelastic scattering is introduced 
and reports given in the session on diffractive interactions 
at the International Workshop on Deep-Inelastic Scattering and Related 
Phenomena, Rome, April 1996, are presented.
}

\section{Introduction}
Since the first observations of large rapidity gap events in deep-inelastic
scattering at HERA
\cite{firstZEUS,firstH1}, diffractive interactions have attracted much attention. 
In this Conference, 31 reports (about half theoretical and half experimental)
have been presented to the working group on 
diffraction, of which 15 were presented in sessions held in common with the 
working groups on structure functions, on photoproduction and on final states.
Two discussion sessions were devoted mainly to the interpretation of
the inclusive measurements.
Reports were also presented on the DESY Workshop on the future of HERA \cite{halina}, 
and on Monte Carlo simulations of diffractive processes \cite{solano}.
The experimental results concern mainly HERA, but also experiments at the 
Tevatron collider.
The present summary consists of two parts, devoted respectively to inclusive 
measurements 
\footnote{presented by V. Del Duca and P. Marage} and to exclusive vector meson 
production \footnote{presented by E. Gallo}.

\section{DDIS Inclusive Measurements}

\subsection {Introduction}\label{sec:intr}

\begin{figure}[h]
 \begin{center}
 \epsfig{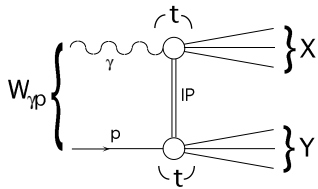}
 \end{center}
  \caption  {Diffractive interaction in photo-- or electroproduction 
              $\gamma^{(*)} p \rightarrow XY$.}
  \label{diffproc}
\end{figure}

Diffractive interactions,
sketched in Fig. \ref{diffproc}, are attributed to the exchange of a colour singlet 
system, the pomeron, and are characterised by the presence in the
final state of a large rapidity gap, without particle emission, between two 
systems \X\ and \Y\ of masses \mx\ and \my\ much smaller than the
total hadronic mass \W.
The final state system \Y\ is a proton (elastic scattering) or an excited 
state of higher mass (proton dissociation).

The cross section for diffractive deep inelastic scattering (DDIS)
is defined using four variables 
(in addition to the mass \my\ of the system \Y), 
$e.g.$ \qsq\ (the negative four-momentum squared of the exchanged
photon), the Bjorken scaling variable \x, the total hadronic mass \W, 
and the square of the pomeron four-momentum $t$. 
It is also useful to define the variables \xpom\ and $\beta$
\begin{equation}
\xpom = \frac{\qsq+\mx^2-t}{\qsq+\wsq-\mpr^2}, \ \ \ \beta = \frac{\qsq}{\qsq+\mx^2-t} ,
    \label{eq:eqPMI}
\end{equation}
which are related to \x\ by the relation $x = \beta \cdot \xpom $. 

Inclusive DIS is usually parameterised in terms of two structure functions,
$F_i = F_i(x, Q^2)$, with $i=1,2$. DDIS needs in principle four
structure functions, which can be written in such a way that two
are similar to the ones of inclusive DIS, and the other two are
the coefficients of terms proportional to the $t$ variable,
and may be neglected since $t$ is small~\cite{chpww}. Thus, in analogy with
inclusive DIS the cross section for DDIS is written as
\be
{d^4\sigma(e + p \ra e + X + Y)
\over dx dQ^2 dx_{I\!\! P} dt} = 
{4\pi \al^2\over x Q^4}\, \left[1-y+{y^2\over 2(1+R^D)}\right]\,
F_2^{D(4)}(x, Q^2; x_{I\!\! P}, t)\, ,\label{nove}
\ee
with $y = \qsq / x \cdot s$ the electron energy loss, $s$ the total 
$(e, p)$ centre of mass energy, and
\be
R^D(x, Q^2; x_{I\!\! P}, t) = {1\over 2x} {F_2^{D(4)}
\over F_1^{D(4)}} - 1\, .\label{ratio}
\ee
$R$ has not been measured yet in the inclusive DIS, 
and $R^D$ will be set to 0 in what follows. Therefore
the diffractive cross section (\ref{nove}) is directly related to 
$F_2^{D(4)}$~\footnote{In the theoretical literature the diffractive 
structure function $F_2^{D(4)}$ is often called 
${dF_2^D \over dx_{I\!\! P} dt}$.}.
If factorization of the collinear 
singularities works in this case as it does in inclusive DIS, then
the diffractive structure function may be written in terms of
a parton density of the pomeron~\cite{ber},
\be
F_2^{D(4)}(x, Q^2; x_{I\!\! P}, t) = \sum_a 
\int^{x_{I\!\! P}}_x d\zeta {df^D_{a/p}(\zeta,\mu; x_{I\!\! P},t)
\over dx_{I\!\! P} dt} \hat F_{2,a}\left({x\over\zeta}, Q^2, \mu\right)\,
,\label{diffact}
\ee
with $\zeta$ the parton momentum fraction within the proton, 
$x_{I\!\! P}$ the momentum fraction of the pomeron, $\mu$ the factorization
scale, and the sum extending over
quarks and gluons; the parton structure functions $\hat F_{2,a}$ are
computable in perturbative QCD. The integral of the diffractive parton 
density over $t$ is the fracture function~\cite{trent}
\be
\int_{-\infty}^0 dt 
{df^D_{a/p}(\zeta,\mu; x_{I\!\! P},t) \over dx_{I\!\! P} dt} =
M_{pp}^a(\zeta,\mu; x_{I\!\! P})\, .\label{frac}
\ee
The structure function $F_2^{D(3)}$ is obtained by integration of 
$F_2^{D(4)}$ over the \ttra\ variable. It is thus related to the
fracture function by
\be
F_2^{D(3)}(x, Q^2; x_{I\!\! P}) = \sum_a \int^{x_{I\!\! P}}_x d\zeta 
M_{pp}^a(\zeta,\mu; x_{I\!\! P}) \hat F_{2,a}\left({x\over\zeta}, Q^2, 
\mu\right)\, .\label{newdiff}
\ee
Next, let us assume that Regge factorization holds for the diffractive
parton density, namely that it can be factorized into a flux
of pomeron within the proton and a parton density within the 
pomeron~\cite{is,dland}: 
\be
{df^D_{a/p}(\zeta,\mu; x_{I\!\! P},t) \over dx_{I\!\! P} dt} =
{f_{p I\!\! P}(x_{I\!\! P},t) \over x_{I\!\! P}}
f_{a/I\!\! P}\left({\zeta\over x_{I\!\! P}},\mu; t\right)\, ,\label{regg} 
\ee
with flux 
\be
f_{p I\!\! P}(x_{I\!\! P},t) = {|\beta_{p I\!\! P}(t)|^2\over 8\pi^2}
x_{I\!\! P}^{1-2\al(t)}\, ,\label{flux}
\ee
the pomeron-proton coupling $\beta_{p I\!\! P}(t)$ and the trajectory
$\al(t)$ \footnote{For simplicity the trajectory is supposed to be 
linear~\cite{dl2}, however it is also possible to consider models
with a non-linear trajectory~\cite{jenk}.} being obtained from fits to elastic 
hadron-hadron cross sections at small $t$ ~\cite{chpww,dl2},
\bea
\beta_{p I\!\! P}(t) &=& \beta_{\bar{p} I\!\! P}(t) \simeq 4.6\, mb^{1/2}\,
e^{1.9 GeV^{-2} t}\, ,\label{five}\\
\al(t) &\simeq& 1.08 + 0.25\, GeV^{-2}\, t\, .\nonumber
\eea
Substituting the diffractive parton density (\ref{regg}) into the
structure function (\ref{diffact}), we obtain
\be
F_2^{D(4)}(x, Q^2; x_{I\!\! P}, t) = f_{p I\!\! P}
(x_{I\!\! P},t)\, F_2^{I\!\! P}(\beta, Q^2; t)\, ,\label{elev}
\ee
with the pomeron structure function
\be
F_2^{I\!\! P}(\beta, Q^2; t) = \sum_a \int_{\beta}^1 d\beta'
f_{a/I\!\! P}\left(\beta',\mu; t\right)
\hat F_{2,a}\left({\beta\over\beta'}, Q^2, \mu\right)\, ,\label{pomf}
\ee
with $\beta' = \zeta/x_{I\!\! P}$ and $\beta$ the fraction 
of the pomeron momentum carried by the struck parton.
When the outgoing proton momentum is measured as the fraction $x_L$ of the
incident proton momentum, one has $x_L \simeq 1 - \xpom$.\\

Several reports at this Conference and numerous discussions dealt with the
procedures of diffractive cross section measurement, the factorisation properties of the
structure function and the possibility to extract parton distributions for the
exchange  system.

\subsection {Cross section measurements}

The H1~\cite{H193} and ZEUS~\cite{ZEUSI} experiments have measured the cross section
for diffractive deep inelastic scattering in the data taken in 1993, by selecting
events with a large rapidity gap in the forward part of their main calorimeters.
The non-diffractive and the proton dissociation diffractive contributions
were subtracted using Monte Carlo simulations.
Within the limited statistics, both experiments found that the results were compatible
with a factorisation of the structure function \fddt\ of the form
\be
\fddt (\qsq, \beta, \xpom) = \frac {1} {\xpom^n} \ A (\qsq, \beta),
   \label{eq:eqPMIII}
\ee
%
where the \xpom~dependence could be interpreted as proportional to a pomeron flux 
in the proton, in agreement with eq.~(\ref{elev}) integrated over $t$.
The exponent $n$ is related to the effective pomeron trajectory 
by $n = 2\ \alpha(t) - 1$, as in eq.~(\ref{flux}),
with $\alpha(t)$ given in eq.~(\ref{five}).

The following \ttra~averaged \albar\ values were obtained by H1 and ZEUS, respectively:
\bea
\albar = 1.10 \pm 0.03 \pm 0.04 \ \ \ {\rm H1} \\
\albar = 1.15 \pm 0.04 \ ^{+0.04}_{-0.07} \ \ \ {\rm ZEUS} . 
\eea

\begin{figure}
\includegraphics{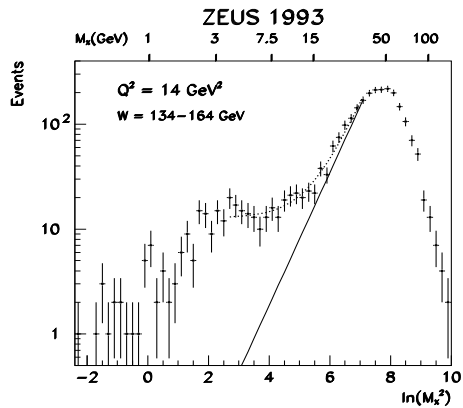}
\unitlength1cm
\begin{picture}(5,5.5)
\thicklines
\end{picture}
\caption{\label{fig:kowalskiI}
             {ZEUS Coll. (log $M_X$ method [14]): 
             Example of a fit for the determination of the nondiffractive
             background.  
             The solid lines
             show the extrapolation of the nondiffractive background as
             determined from the fit of the diffractive and nondiffractive
             components to the data (dotted line).
             }}
\end{figure}

\begin{figure}
\includegraphics{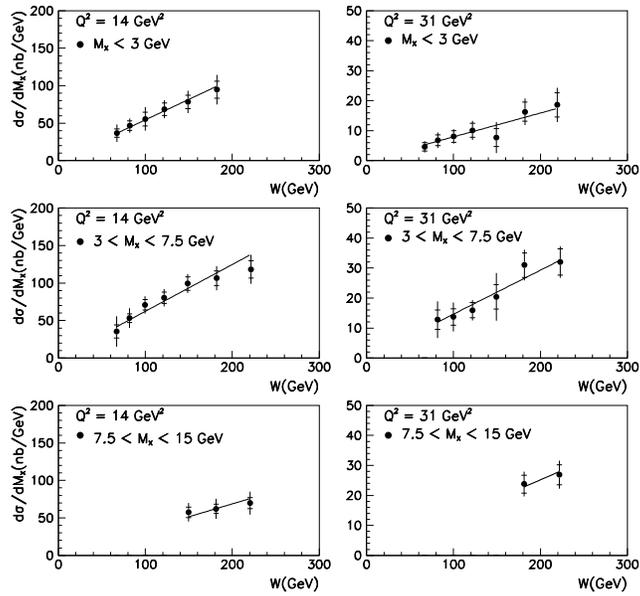}
\unitlength1cm
\begin{picture}(7,8.8)
\thicklines
\end{picture}
\caption{\label{fig:kowalskiII}
            {ZEUS Coll. (log $M_X$ method [14]): 
             The differential cross sections $d\sigma^{diff}
             (\gamma^* p \to X N)/dM_X$. The inner error bars show the 
             statistical errors and the full bars the statistical and
             systematic errors added in quadrature. 
             The curves show the results 
             from fitting all cross sections to the form  $d\sigma^{diff}/dM_X 
             \propto (W^2)^{(2\overline{\alphapom}-2)}$ with a common value of 
             $\overline{\alphapom}$.}}
\end{figure}
The ZEUS Collaboration has presented at this Conference a different method to 
extract the diffractive contribution (1993 data)~\cite{ZEUSII}. 
The (uncorrected) $\log \mx^2$ distributions, in bins of \qsq\ and $W$, are 
parameterised as the sum of an exponentially falling contribution at high \mx, 
attributed to non-diffractive interactions,
and of a constant contribution at low \mx, attributed to diffraction.
(see Fig. \ref{fig:kowalskiI}). 
With this ``operational definition'' of diffraction (with $\my\ \lsim\ 4$ \gevcsq), 
no Monte Carlo simulations are used. 
The $W$ dependence of the diffractive cross section gives
(see Fig. \ref{fig:kowalskiII}):
%
\be
\albar = 1.23 \pm 0.02 \pm 0.04, \ \ \beta = 0.1 - 0.8. 
\ee
The difference with the previously published result is attributed by ZEUS to the 
uncertainties in the Monte Carlo procedure for background subtraction.

\begin{figure}
\vspace{-2cm}
\begin{center}
\leavevmode
\hbox{%
\epsfxsize = 3.5in
\epsffile{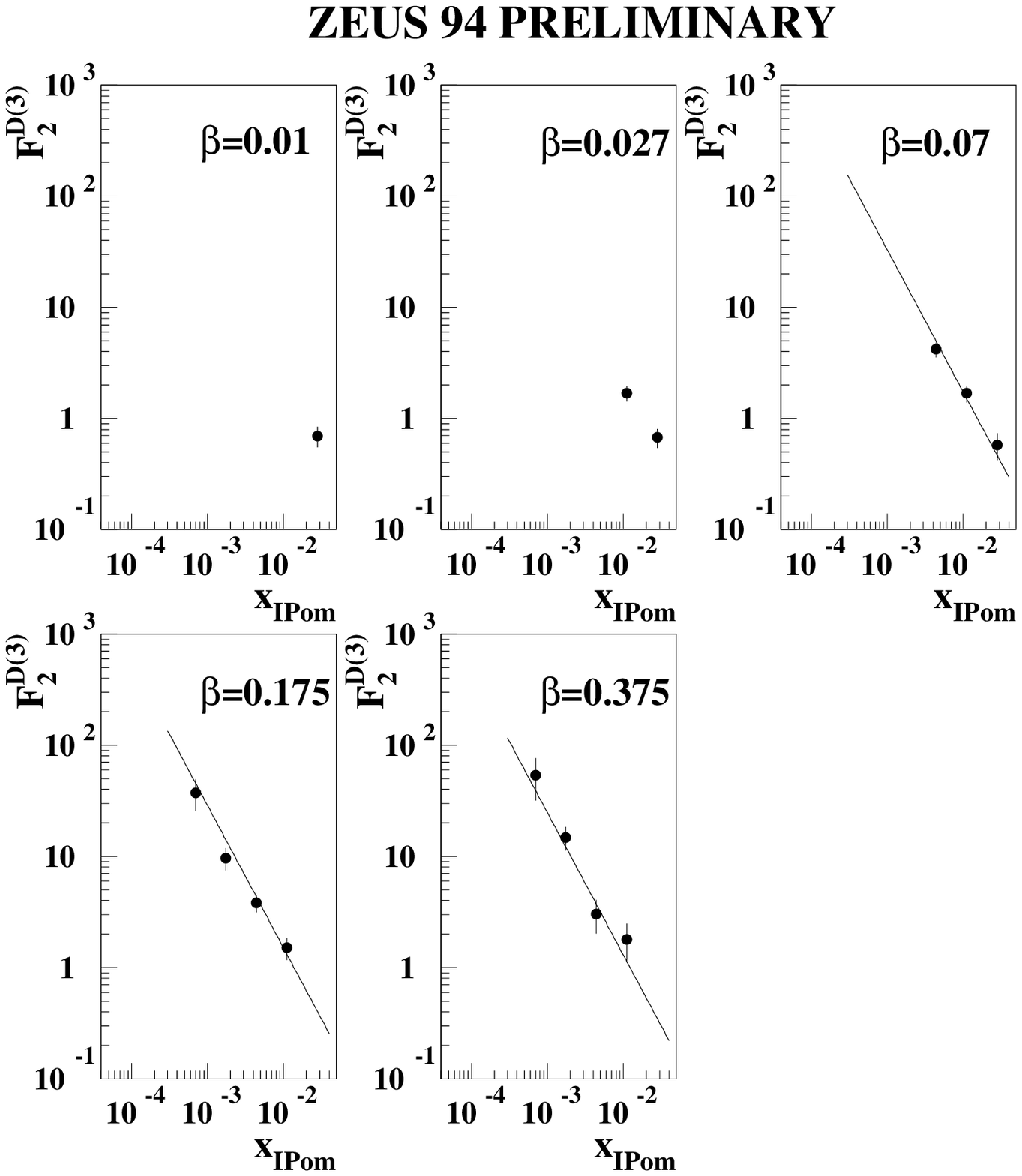}}
\end{center}
\vspace{-2.5cm}
\caption 
{ZEUS Coll. (LPS data [15]): 
The structure function $F^{D(3)}_{2}$, plotted as a function
of $x_{I\!P}$ in 5 bins in $\beta$ at a central value $Q^2=12\ \gevsq$. 
The errors are statistical only. 
The solid line corresponds to a fit in the form of eq. (12).
\label{fig:barberis}}
\end{figure}

The ZEUS Collaboration has also presented in this Conference preliminary results 
obtained with the Leading Proton Spectrometer (1994 data)~\cite{barberis}.
In this case, the scattered proton is unambiguously tagged, and the kinematics are
reconstructed using its momentum measurement. A fit of the \xpom\ dependence
(see Fig. \ref{fig:barberis})
for $\av{\mx^2} = 100$ \gevsq\ gives:
\be
\albar = 1.14 \pm 0.04 \pm 0.08, \ \ \beta = 0.04 - 0.5. 
\ee

The ZEUS LPS has also provided the first inclusive measurement of the diffractive 
$t$ dependence at HERA, parameterised as 
\be
\frac {{\rm d}\sigma} {{\rm d}t} \propto e^{-b|t|}, 
       \ \ b = 5.9 \pm 1.3 \ ^{+1.1}_{-0.7}\ \gevsqm.
\ee

\begin{figure}[ht]
 \begin{center}      
 \epsfig{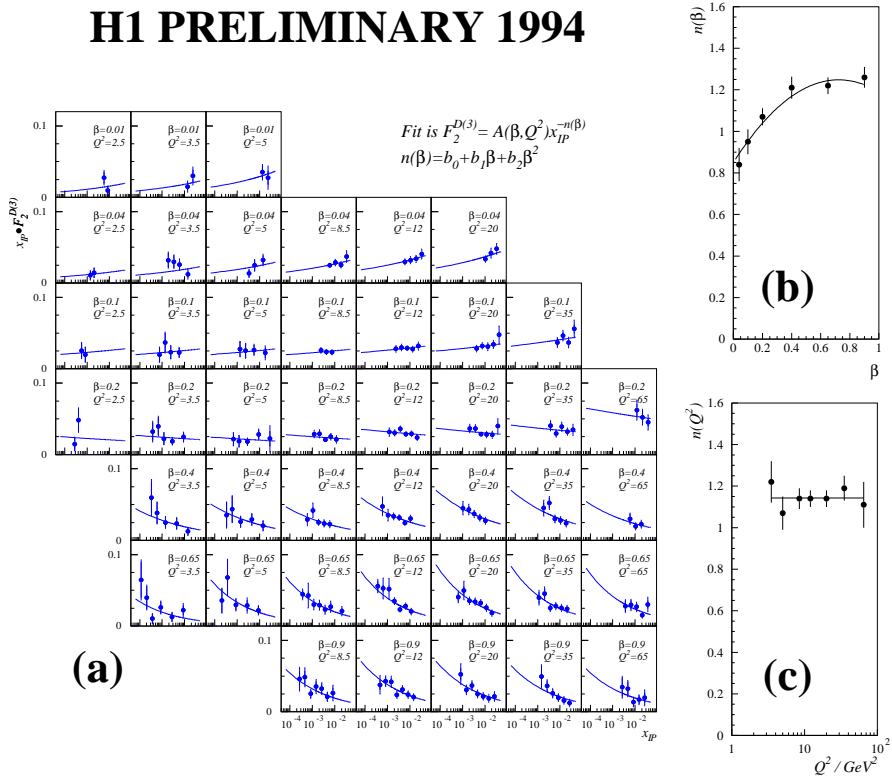}
 \vspace{-0.8cm}
 \caption{H1 Coll. [16]: (a) \protect{$\xpom \cdot F_2^{D(3)}(\beta,Q^2,\xpom)$} integrated
over the range $\my < 1.6\ {\rm GeV}$ and $|t|\,<\,1$ GeV$^2$; (b) the 
$\beta$ dependence of $n$ when $F_2^{D(3)}$ is fitted to the form 
\protect{$F_2^{D(3)}=  A(\beta,Q^2)/ {\xpom}^{n(\beta)}$}; (c) the $Q^2$
dependence of $n$ when $F_2^{D(3)}$ is fitted to the form 
\protect{$F_2^{D(3)}=  A(\beta,Q^2)/ {\xpom}^{n(Q^2)}$}.
The experimental errors are statistical and systematic added in quadrature.}
\label{fig:newmanI}   
\end{center}
\end{figure}

Finally, the H1 Collaboration has reported preliminary results on \fddt\ from
the 1994 data~\cite{newman}. 
The use of the forward detectors allows the selection of events with no activity 
in the pseudorapidity range $3.2 < \eta < 7.5$ ($\my < 1.6$ \gevcsq).
With this extended kinematical domain and a tenfold increase in statistics
compared to the 1993 data, 43 bins in \qsq\ and $\beta$ are defined.
A clear breaking of factorisation is observed: 
in the form of parameterisation (\ref{eq:eqPMIII}), the data suggest that 
the $n$ exponent is independent of \qsq\, but they require a definite $\beta$ dependence,
with \albar\ ranging from $\simeq 0.98$ for $\av{\beta} \simeq 0.1$ to
$\simeq 1.12$ for $\av{\beta} \ \gsim\ 0.4$ 
(see Fig. \ref{fig:newmanI}).

Experimentally, for similar values of $\av{\beta}$, the H1 results thus favour a 
smaller value of \albar\ than the ($\log \mx^2$) ZEUS analysis. 
Detailed comparisons and discussions between the two experiments should in the
future provide more information concerning the source of this difference.

\subsection {Factorisation Breaking and Parton Distributions}
\label{sec:break}

The source of the factorisation breaking observed by H1 has been discussed in 
several communications and during the round tables.

N. N. Nikolaev underlined particularly that the pomeron is not a particle, and that
in a QCD inspired approach, factorisation is not expected to hold~\cite{nikolaev}.

The possible contribution in the selected samples of different exchanges, 
in particular of pomeron and $f$ and $a_2$ trajectories, was particularly
emphasised~\cite{jenk,nikolaev,landshoff,Eilat,pred,stirling}. 
These trajectories have different energy and thus $\xpom^{n}$ dependences. 
They have also different partonic contents, and thus different functions 
$A (\qsq, \beta)$ describe the interaction with the photon. 
Even if each contribution were factorisable, their combination would thus not be 
expected to allow a factorisable effective parameterisation.\\

However, 
if it is possible to select a domain where pomeron exchange dominates, and
if the factorization picture outlined in sect.~\ref{sec:intr} holds,
it is possible to fit the data on single hard diffraction to extract the
parton densities in the pomeron~\footnote{It is not clear whether the fits 
should be extended to data from hadron-hadron scattering or from 
photoproduction because of additional 
factorization-breaking contributions~\cite{cfs}.}.
First we note that the pomeron, being an object with the quantum numbers of
the vacuum, has $C = 1$ and is isoscalar~\cite{dl2}~\footnote{A 
$C =-1$ contribution would be indication of an odderon exchange~\cite{pred}.}. 
The former property implies that
$f_{q/I\!\! P}(\beta) = f_{\bar{q}/I\!\! P}(\beta)$ for any quark $q$
and the latter that $f_{u/I\!\! P}(\beta) = f_{d/I\!\! P}(\beta)$. 
Therefore it is necessary to determine only the up and strange quark densities
and the gluon density. Since the parton structure function is,
\be
\hat F_{2,a}\left({\beta\over\beta'}, Q^2, \mu\right) = e_a^2 
\delta\left(1 - {\beta\over\beta'}\right) + O(\al_s)\, ,\label{partf}
\ee
with $e_a$ the quark charge and $a$ running over the quark flavors,
the pomeron structure function (\ref{pomf}) becomes,
\be
F_2^{I\!\! P}(\beta, Q^2; t) = {10\over 9} \beta f_{u/I\!\! P}(\beta,Q^2; t) +
{2\over 9} \beta f_{s/I\!\! P}(\beta,Q^2; t) + O(\al_s)\, ,\label{dod}
\ee
where the gluon density contributes to the $O(\al_s)$ term through the
DGLAP evolution.\\

\begin{figure}[ht]
\begin{center}
  \epsfig{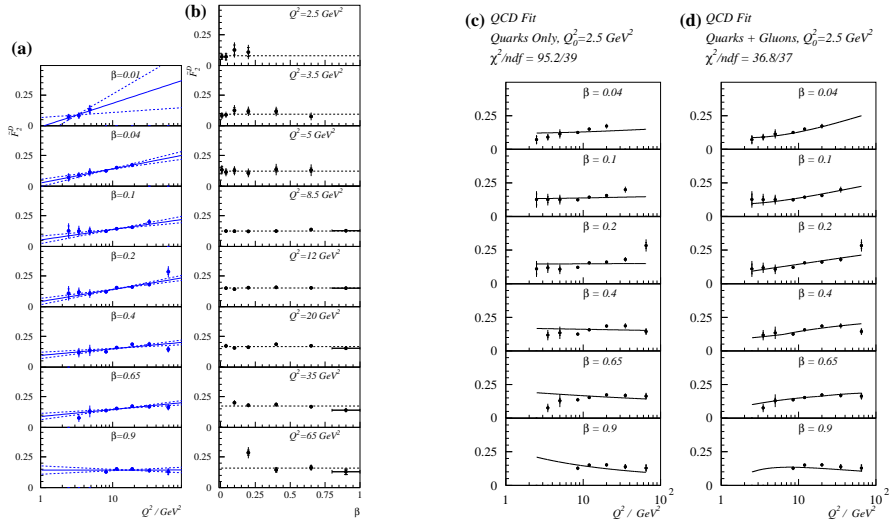} 
   \end{center}
 \vspace{-0.4cm}
 \caption{H1 Coll. [16]: (a) \protect{${\tilde{F}}_{2}^{D}(\beta,Q^2)$} as 
   a function of $Q^2$ for different $\beta$ values. The superimposed
 lines correspond to the best fit to a linear dependence on $\ln Q^2$ 
(continuous) and $\pm 1\sigma$ (dashed); 
(b) \protect{${\tilde{F}}_{2}^{D}(\beta,Q^2)$} as a function of $\beta$ for 
different $Q^2$ values, with a best fit to a constant $\beta$ dependence;
(c) DGLAP QCD comparison of the $(\beta,Q^2)$ dependence of
${\tilde{F}}_{2}^{D}$ assuming
only quarks at the starting scale of $Q_0^2=2.5\,{\rm GeV^2}$; (d) DGLAP QCD
comparison of the $(\beta,Q^2)$ dependence of ${\tilde{F}}_{2}^{D}$ assuming
both quarks and gluons at the starting scale.}
\label{fig:newmanII}
\end{figure}

The H1 Collaboration has studied the evolution of the structure function
\fddt, integrated over \xpom\ in the range $0.0003 < \xpom\ < 0.05$ (in practice,
most of the data are for $\xpom\ < 0.02$):
\be 
\fddtild(\qsq,\beta) =  \int \fddt(\qsq,\beta,\xpom) \ {\rm d}\xpom.
\ee
It is observed 
(see Fig. \ref{fig:newmanII}) 
that $\fddtild(\qsq,\beta$) shows
no $\beta$ dependence at fixed \qsq\, but increases with \qsq\ for fixed $\beta$ values,
up to large $\beta$. 
If interpreted in a partonic framework, this behaviour, strinkingly different of that
of ordinary hadrons, is suggestive of an important gluonic contribution at large
$\beta$.

More specifically, the H1 Collaboration assumed the possibility to perform 
a QCD analysis of this evolution of \fddtild\ using the DGLAP equations (with
no inhomogeneous term) to extract parton densities in the exchange.
At $\qsq = 5\ \gevsq$, a leading gluon component is obtained.
When the corresponding parton densities are input in Monte Carlo simulations of
exclusive processes (sect.~\ref{sec:vddone}), 
consistent results are obtained \cite{theis,tap,alice}.

This procedure was discussed during the round tables and in several 
contributions.
In the absence of factorisation theorems (see ~\cite{berera}), it can be 
questioned whether the parton distribution functions are universal, and whether
they obey a DGLAP evolution. 
A specific problem, due to the fact that the pomeron is not a particle, is that
momentum sum rules need not be valid.
However, it was noticed that the contribution of several Regge trajectories 
may not affect the validity of a common QCD-DGLAP evolution, in so far as these 
exchanges can all be given a partonic interpretation~\cite{stirling}.

On the other hand, it was also argued~\cite{nikolaev,bartels} that, even if one
accepts the concept of parton density functions in the diffractive exchange,
the DGLAP evolution should not be valid at high $\beta$, because of charm
threshold effects and because of the specific and different \qsq\ evolutions
of the longitudinal and transverse contributions.

\subsection{Parton Distributions and Jet Production}
\label{sec:vddone}

Jet production in diffractive interactions has yielded the first
hint of a partonic structure of the pomeron~\cite{is}. We have seen
in sect.~\ref{sec:break} how the quark densities may be directly
related to the pomeron structure function.
The gluon density may also be directly measured by using data
on diffractive charm or jet production. For the latter,
the final state of the hard scattering $jet_1 + jet_2 + X$ consists 
at the lowest order $O(\al_s)$ of two partons only, generated in
quark-exchange and Compton-scattering diagrams for quark-initiated
hard processes, and in photon-gluon fusion diagrams for the gluon-initiated
ones. The parton momentum fraction $x$ in the proton may be 
computed from the jet kinematic variables, $x = (E/P^0) \exp(2\bar{\eta})$,
with $E$ and $P^0$ the electron and proton energies and $\bar{\eta} = 
(\eta_{j_1}+\eta_{j_2})/2$ the rapidity boost of the jet system. The 
momentum fraction $x_{I\!\! P}$ may be obtained
from the invariant mass of the
system recoiling against the proton, $x_{I\!\! P}\simeq M_{ej_1j_2}^2/s$.
If we neglect the strange quark density~\footnote{The strange quark 
density might be measured adding to the fit data on charged-current
charm production in DDIS~\cite{chpww}.}, the data on $F_2^D$ and diffractive
jet production suffice to measure the parton densities in the pomeron, and 
may be used to link the value of the momentum sum, 
$\sum_a \int_0^1 dx\, x f_{a/I\!\! P}(x)$, to the gluon content of the
pomeron~\cite{msr}. 

On the other hand, if the gluon density is determined from $F_2^D$ alone,
as outlined in sect.~\ref{sec:break}, one can
make predictions for the data on diffractive charm or jet production.
Indeed, using the data on $F_2^D$ and the Monte Carlo RAPGAP~\cite{jung}, 
based on the factorization (\ref{elev}), the H1 Collaboration~\cite{newman} 
finds that the parton densities $f_{a/I\!\! P}$ are dominated by a very
hard gluon~\footnote{There are models~\cite{cfs,buch} which predicted the 
dominance of a single
hard gluon. In these models the color is neutralized by the exchange of
one or more soft gluons.} (sect.~\ref{sec:break}). 
Having measured the densities $f_{a/I\!\! P}$, the H1 
Collaboration~\cite{theis} finds that the jet rapidity and
transverse momentum distributions in diffractive dijet production 
are in good agreement with the prediction from RAPGAP. In addition, 
the data on energy flow in diffractive photoproduction also
seem to support a gluon-dominated structure of the pomeron~\cite{tap}.

Besides, by examining the thrust we may probe the amount of
gluon radiation in diffractive dijet production. The thrust axis is 
defined as the axis in the parton center-of-mass system
along which the energy flow is maximal. The value of the thrust, $T$, 
then measures the fraction of energy along this axis. For a back-to-back
dijet event $T=1$; so the amount by which $T$ differs from unity gives an
indication on the amount of emitted gluon radiation.
The data~\cite{alice} shows indeed the presence of hard gluon radiation,
the thrust being even smaller than in dijet production in $e^+e^-$
annihilation.

Finally, we mention several angular-correlation analyses recently proposed.
Namely, the
azimuthal-angle distribution in diffractive dijet production~\cite{bartels};
the final-state electron-proton azimuthal-angle distribution in the lab
frame~\cite{stir}; the azimuthal-angle distribution of the initial electron
in the photon-proton frame~\cite{nacht}. The respective measurements, if
carried out, should allow to further probe the pomeron structure, and to
discriminate between different models.

\subsection {Other Measurements}

In addition to these diffractive DIS measurements, the HERA Collaborations
also contributed reports on several other topics.

The H1 Collaboration presented results on diffraction in 
photoproduction~\cite{newman}, with a measurement of its decomposition in
vector meson production, photon dissociation, proton dissociation and
double dissociation. In particular, the \mx\ distribution is consistent
with soft pomeron exchange.

The ZEUS experiment reported the observation of a charged current diffractive
candidate event in the 1994 data~\cite{zarnecki}, and discussed the design of a 
special trigger used in the 1995 data taking.

ZEUS also presented the observation of DIS events with a high energy neutron
produced at very small angle with respect to the proton direction, 
detected in their neutron counter~\cite{jmartin}.
These events, attributed to pion exchange, account for about 10\% of the
total DIS rate, independent of $x$ and \qsq.

Finally, the E665 muon experiment at Fermilab reported on the ratio of diffractive
to inelastic scattering~\cite{Wittek}.


\section{Exclusive Vector Meson Production}

\subsection{Introduction}\label{intro:vm}

Exclusive vector mesons production at HERA\footnote{Combined session
with Working Group 2, on Photoproduction Interactions.}
 is a very interesting process to study
the transition from a non-perturbative description of the pomeron, the 'soft'
pomeron,  to the hard perturbative pomeron.

The process that we study 
is shown in figure \ref{fig1:vm}a and corresponds to the reaction
\begin{equation}
ep \rightarrow e V N,
\label{eq1:vm}
\end{equation}
where $V$ is a vector meson ($\rho,\omega,\phi,\rho^\prime,J/\psi,...$) and $N$ is either
the final state proton which remains intact in the interaction (elastic production)
or an excited state in case of proton dissociation.

The cross section for the elastic process has been calculated by several authors. In the
'soft' pomeron picture of Donnachie-Landshoff \cite{dl}, 
the photon fluctuates into a $q \bar q$ pair,
which then interacts with the proton by exchanging a pomeron.
The cross section $\sigma(\gamma p \ra Vp)$ is expected to increase slowly with the $\gamma p$
center of mass energy $W$; the exponential dependence on $t$, the  square of the
four momentum transfer at the proton vertex, is expected to become steeper as $W$ increases
(shrinkage). 

In models \cite{ryskin,brodsky,nemchik} based on perturbative QCD,
the pomeron is treated as a perturbative two-gluon system: the cross section is then related to the
square of the gluon density in the proton and  a strong dependence of the 
cross section on $W$ is expected. The prediction,
taking into account  the HERA measurements of the gluon density at
low $x_{\rm Bjorken}$, is that the cross section should have 
at low $x$ a dependence of the type $W^{0.8-0.9}$.
 In order to be able to apply perturbative QCD, a 'hard' scale
has to be present in the process. 
The scales involved in process (\ref{eq1:vm}) are the mass of the quarks in the vector meson $V$,
the photon virtuality $Q^2$ and $t$. 

In the following we 
summarize results on vector meson production obtained by the H1 and ZEUS Collaborations
at HERA, in the energy range $W =40-150 ~{\rm GeV}$,
starting from $Q^2 \simeq 0$ and increasing the scales in the process.
We also review 
results at high $t$, which  were presented for the first time at this conference. 
Results on vector meson production with diffractive proton dissociation events were also discussed.  


\subsection{Vector mesons at $Q^2 \simeq 0$}\label{mass:vm}

\begin{center}
\begin{figure}
\hbox{
\psfig{figure=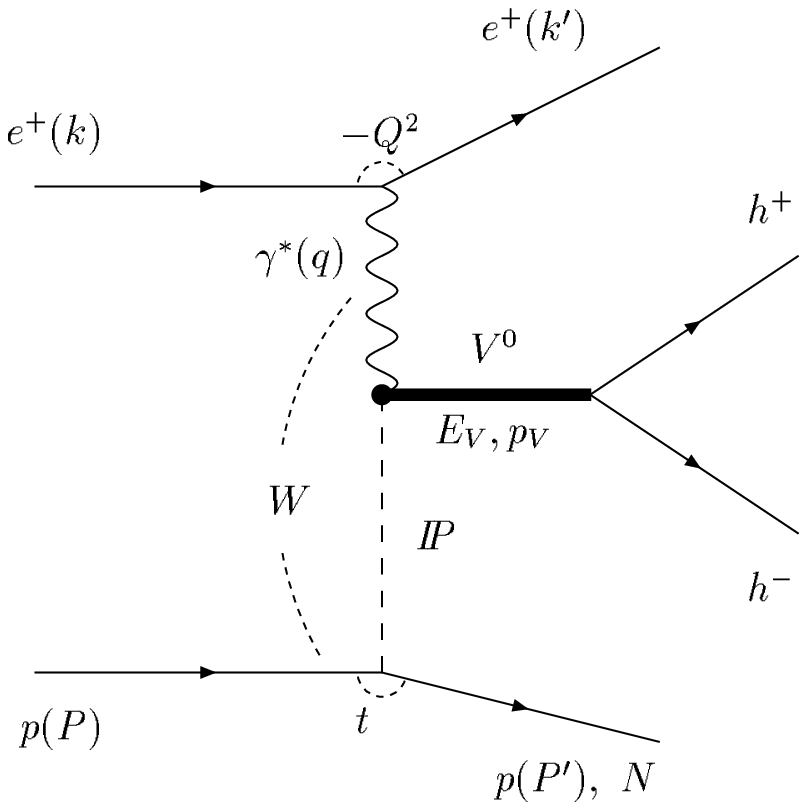,height=2.0in}
\psfig{figure=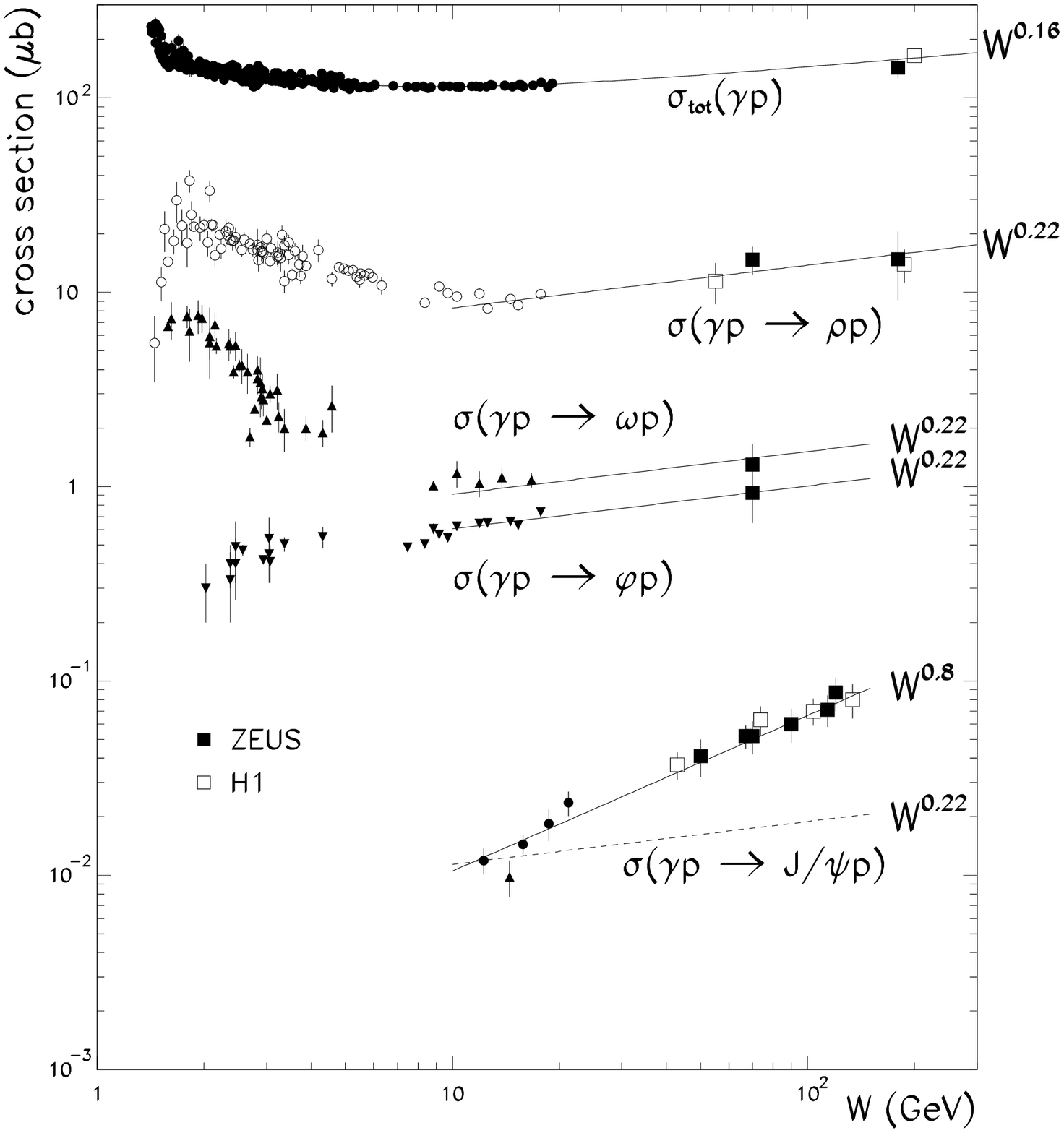,height=2.5in}}
\caption{(a) Feynman graph of vector meson production at HERA;
(b) Cross section versus $W$ for vector meson production at $Q^2 \simeq 0, t \simeq 0$.}
\label{fig1:vm}
\end{figure}
\end{center}

Elastic vector meson  production at $Q^2 \simeq 0$ has been studied by both Collaborations
\cite{sacchi,schiek}. The cross section for $\rho,\omega,\phi,J/\psi$ production
is plotted versus $W$
in fig.~\ref{fig1:vm}b (from \cite{h1rho}, where the results obtained at HERA  ($W=50-200$ GeV) 
are compared to those 
of fixed target experiments  ($W \simeq 10$ GeV).
At $Q^2 \simeq 0$, the cross section is mainly due to transversely polarized photons.
The Donnachie-Landshoff model (DL), where one expects for the 
$\sigma(\gamma p) \ra Vp$ cross section a dependence  
of the type $\simeq W^{0.22}$,
reproduces the energy dependence for the light vector mesons (see lines in the figure).
In the same way the dependence of the total photoproduction cross section  is well reproduced
by the soft pomeron model \cite{landshoff} (see figure). 
In contrast, this model fails to describe the strong $W$ dependence observed
in the $J/\psi$ data. Note that this steep $W$ dependence,  
$\sigma(\gamma p \ra Vp) \simeq W^{0.8}$, is
implied even within the restricted $W$ range covered by the HERA data alone. 
In this case the hard scale is
provided by the charm mass.

Elastic $J/\psi$ production is a very important process to determine the gluon density in the
proton, as the cross section is proportional to the square of this density. 
The gluon density can be measured in the range
$5 \times 10^{-4} < x \simeq \frac{(m_V^2+Q^2+|t|)}{W^2} < 5 \times 10^{-3}$.
 An improved  calculation for this process was  presented \cite{amartin}: 
although the normalization of
the cross section is known theoretically with a precision  of $30\%$, the shape of the $W$ dependence is very
sensitive to different parton density  parametrizations in the proton.
  Open charm production is also a very sensitive  probe
of the gluon density, and the perturbative calculation has no ambiguities in the normalization; however it is
experimentally more difficult.

\subsection{Vector mesons at high $Q^2$}\label{qtwo:vm}

\begin{figure}
\hbox{
\psfig{figure=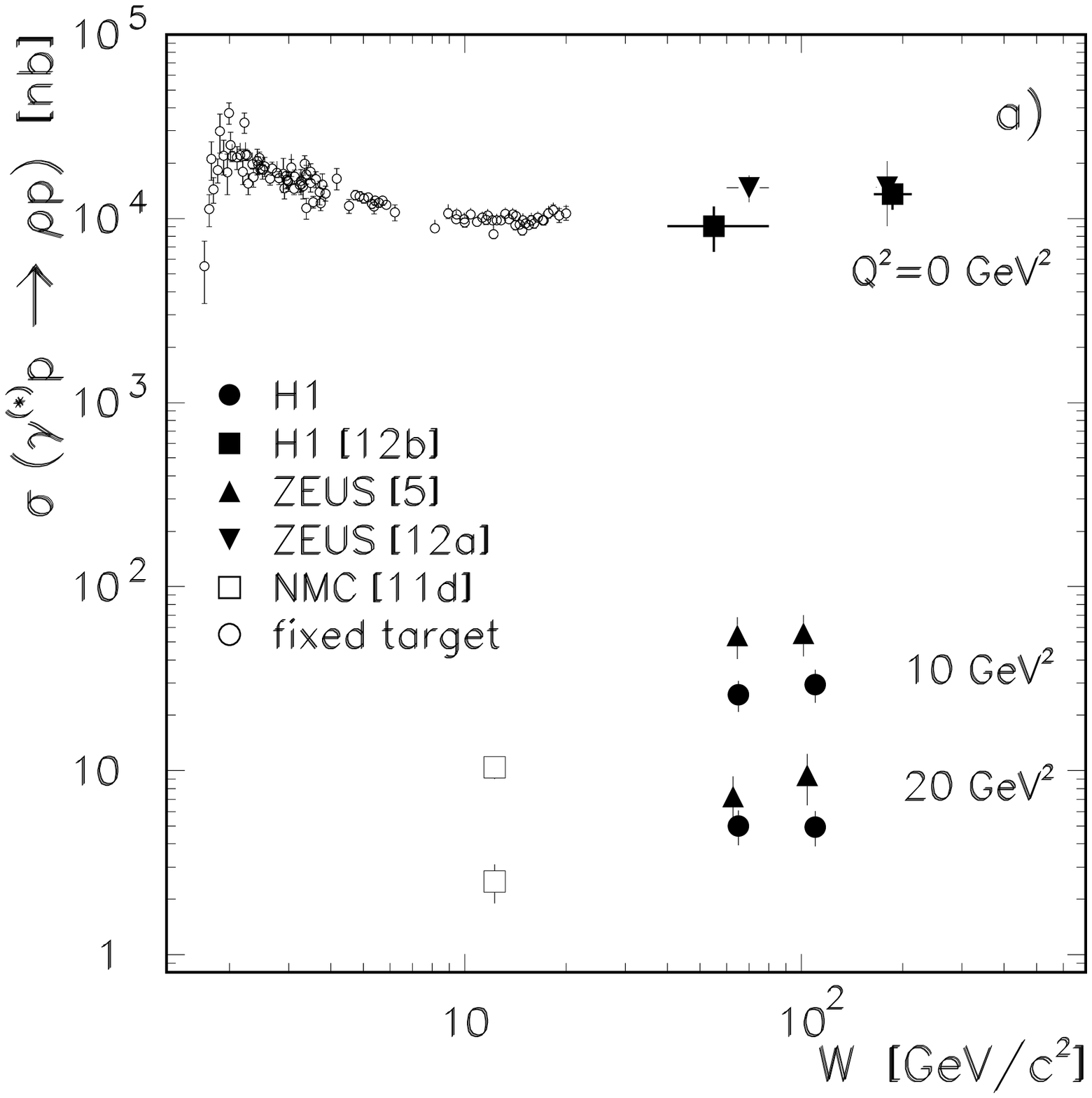,height=2in}
\psfig{figure=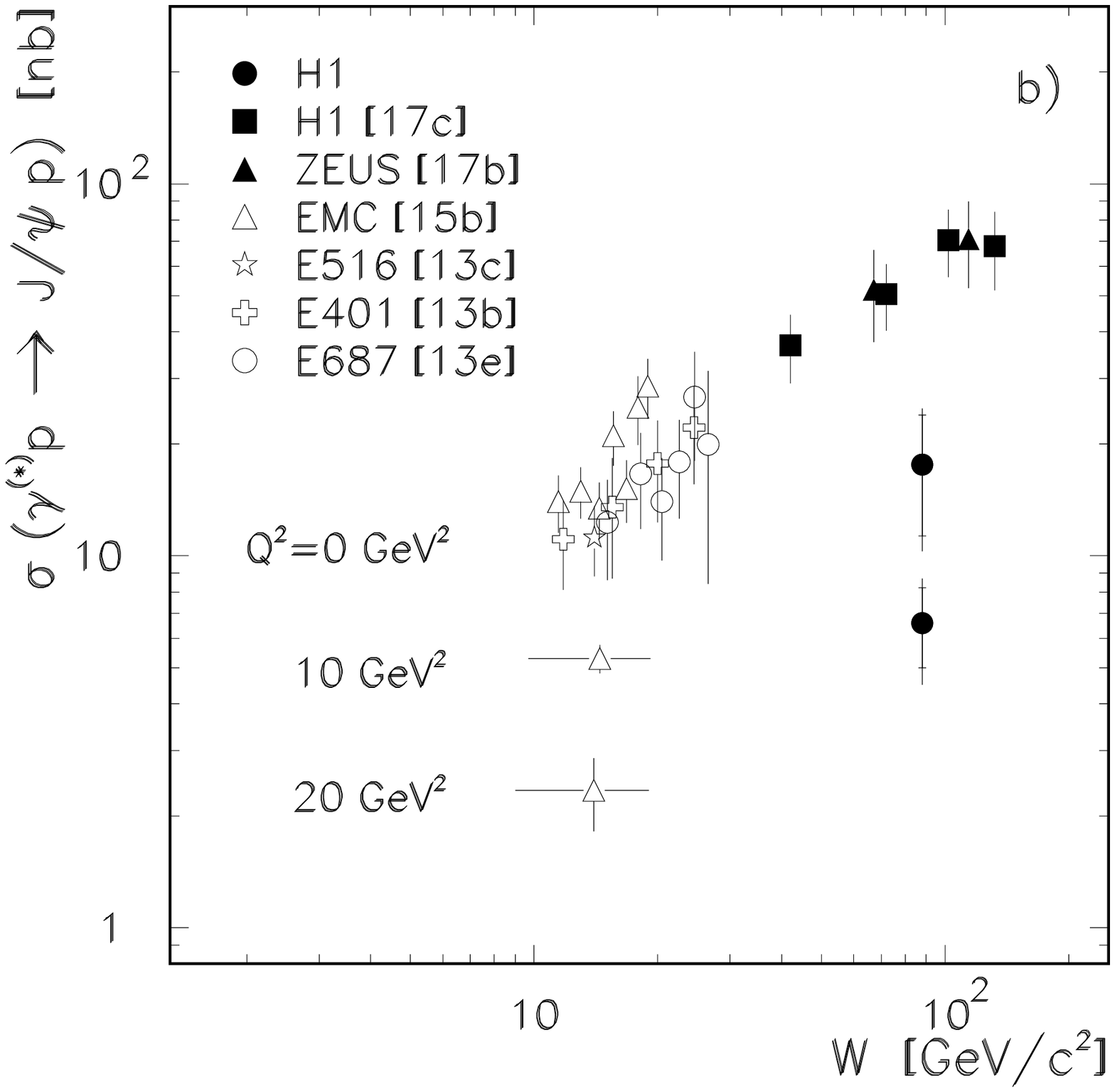,height=2in}}
\caption{Cross section versus $W$ for vector meson production at $Q^2 \simeq 0$ and at high $Q^2$, for $\rho$ (a) and $J/\Psi$ (b) production.}
\label{fig2:vm}
\end{figure}

Results on vector meson production at high $Q^2$ have been presented by H1 \cite{clerbaux}.
 The cross section for the process $(\gamma^*p \ra \rho p)$ with the H1
1994 data is shown in fig.~\ref{fig2:vm}a, together with the ZEUS 1993 results \cite{zeusrho}, 
in the  $\gamma^* p$ centre of mass energy $W$ of $40$ to $140$ GeV, and at 
$Q^2=10,20~{\rm GeV^2}$.
At these values of $Q^2$, the $\sigma(\gamma^*p \ra Vp)$ cross section is dominated by 
longitudinally polarized photons.
Comparing the H1 data to the NMC data at  
$W \simeq 10~{\rm GeV}$, a dependence of the type $\sigma \simeq W^{0.6}$ 
at $Q^2=10~{\rm GeV^2}$ 
for the the cross section is obtained.
The $t$ dependence of the reaction $\gamma^* p \ra \rho p$ 
for $\modt < 0.5 ~{\rm GeV^2}$ is well reproduced by an     
an exponential distribution ${\rm exp}(-b\modt)$, with $b \simeq 7~{\rm GeV^{-2}}$ 
(see table \ref{tab:vm}). 
In the framework of Regge theory, the shrinkage of the elastic peak can be
written as $b(W^2)= b(W^2=W_0^2)+ 2\alpha^\prime \ln (W^2/W^2_0)$, where 
$\alpha^\prime$ is the slope of the pomeron trajectory.
Comparing the H1 1994 data with
the NMC data, the parameter $\alpha^\prime$ 
which is obtained is in agreement with that expected from a soft
pomeron trajectory  ($\alpha^{\prime}=0.25$ \gevsqm).
The H1 1994 and the ZEUS 1993 data are compatible within the errors, however
while the H1  $\rho$ data suggest that we are in a transition region between soft and hard
processes, the ZEUS 1993 data show
a stronger $W$ dependence for the cross section when compared to NMC ($\sigma \simeq W^{0.8}$),
and a flatter $t$ distribution, $ b \simeq 5~{\rm GeV^{-2}}$ (tab. \ref{tab:vm}). 
These two last results suggest a hard pomeron exchange. 
More data will allow to
reduce the uncertainties due to the non resonant background, to the proton dissociation background and to study
the $W$ dependence in the region covered by the HERA data alone.

The ratio of the dissociative over elastic $\rho$ cross section was measured \cite{clerbaux} 
to be $0.59 \pm 0.12 \pm 0.12$, with no significant $Q^2$ dependence (in the range
between $8$ and $36~\rm{GeV^2}$) or $W$ dependence (in the range between $60$ and
$180~\rm{GeV}$). 

$J/\psi$ production was presented by both collaborations \cite{clerbaux,stanco}:  while the ratio of the cross sections
$\sigma(J/\Psi)/\sigma(\rho)$ is of the order of $10^{-3}-10^{-2}$ at $Q^2 \simeq 0$, 
this ratio becomes close to 1 at $Q^2>10~{\rm GeV^{2}}$ (see figs  \ref{fig2:vm}a 
and  \ref{fig2:vm}b, from \cite{h1rho}), as predicted
by perturbative QCD \cite{strikman}.

\subsection{Vector mesons at high $\modt$}\label{t:vm}

The installation in 1995 of a new detector \cite{piotr} 
in the ZEUS experiment at $44~{\rm m}$ from the interaction point in the direction of the outgoing positron,
allowed to tag the scattered positron and to access vector mesons at $Q^2 < 0.01 ~{\rm GeV^2}$, 
$1 < |t| < 4.5 ~{\rm GeV^{2}}$ and in the $W$ range
$80 < W< 100 ~{\rm GeV}$. About 600 $\rho$ candidates and 80 $\phi$ candidates were found:
from Monte Carlo studies it was seen that at $t>1~{\rm GeV}^2$, the main contribution is 
proton dissociative production.
The ratio of the cross sections $\sigma(\phi)/\sigma(\rho)$ was measured in two ranges of $t$ 
($ 1.2 < |t| < 2 ~{\rm GeV^2}$, $ 2 < |t| < 4.5 ~{\rm GeV^2})$, and was found to be 
$0.16\pm 0.025(stat.) \pm 0.02 (sys.)$ and $0.18 \pm 0.05 (stat.) \pm 0.04(sys.)$ respectively,
close to the value of $2/9$ predicted by SU(3) flavour symmetry. This value is significantly higher
than the value obtained at $t \simeq 0, ~Q^2 \simeq 0$ and is instead compatible with the value obtained \cite{clerbaux,zeusphi} at
$Q^2 \simeq 12 ~{\rm GeV^2}$, 
suggesting that at high $|t|$ perturbative QCD plays an important role.
Vector meson production at high $\modt$ has been suggested  \cite{ivanov}
as a very nice field to study the hard pomeron,
since the  cross sections are calculable in pQCD.

\begin{table}[t]
\caption{Results on the value $b$ (in ${\rm GeV}^{-2}$) fitted to the exponential $t$ distributions
for elastic and proton dissociation vector meson production at HERA. The first error is the statistical, the second error is the systematic one.
\label{tab:vm}}
\vspace{0.4cm}

\footnotesize
\begin{center}
\begin{tabular}{|c|c|c|c|}
\hline
                    &        & $b$(H1)      & $b$(ZEUS) \\ \hline
$Q^2=0,|t|<0.5~{\rm GeV}^2$, elastic & $\rho$ & $10.9 \pm 2.4 \pm 1.1$  &  $9.9 \pm 1.2\pm 1.4$ \\
$Q^2=0,0.07<|t|<0.4~{\rm GeV}^2$                     &        &                           & $9.6\pm0.8\pm1.2$ (LPS) \\ \hline  
$Q^2=0,|t|<0.5~{\rm GeV}^2$, elastic & $\omega$ &                     & $10.6\pm 1.1\pm 1.4$ \\ 
$Q^2=0,|t|<0.5~{\rm GeV}^2$, elastic & $\phi$ &                         & $7.3\pm1.0\pm0.8$ \\ 
$Q^2=0,|t|<1~{\rm GeV}^2$, elastic & $J/\Psi$ & $4.0\pm 0.2\pm 0.2$ &                                \\ \hline  
$Q^2\simeq 10,|t|<0.5~{\rm GeV}^2$, elastic & $\rho$ & $7.0 \pm 0.8 \pm 0.6$   &  $5.1^{+1.2}_{-0.9} \pm 1.0$  \\
$Q^2\simeq 10,|t|<0.8~{\rm GeV}^2$, elastic & $J/\Psi$ & $b=3.8 \pm 1.2^{+2.0}_{-1.6}$  &                    \\ \hline
$Q^2=0,0.04<|t|<0.45~{\rm GeV}^2$, p-dissociation & $\rho$ &                              & $5.3 \pm 0.8\pm 1.1$ (LPS)  \\  
$Q^2=10,|t|<0.8~{\rm GeV}^2$, p-dissociation & $\rho$ &   $2.1 \pm 0.7 \pm 0.4 $  &                               \\  
$Q^2=10,|t|<1~{\rm GeV}^2$, p-dissociation & $J/\Psi$ & $1.6 \pm 0.3 \pm 0.1 $ &                                \\ \hline
$Q^2=0,1<|t|<4.5~{\rm GeV}^2$ , p-dissociation & $\rho$ &         & $ \simeq2.5 $ \\ \hline
\end{tabular}
\end{center}
\end{table}
\normalsize

\subsection{Outlook}

A summary of the $t$ slopes presented at this conference is given in table \ref{tab:vm}.
The results marked with LPS were obtained using  the ZEUS  Leading Proton Spectrometer 
\cite{sacchi}, which detects the scattered proton and measures its momentum; the LPS allows
to tag a clean sample of elastic events and to measure $t$ directly.
The parameter $b$ is proportional, in diffractive processes, to the square of the radius of the
interaction, which decreases with the mass of the quark or the photon virtuality, as confirmed by the
results in the table.
Note that the result in inclusive diffractive deep inelastic events obtained by the ZEUS experiment using the
LPS is $b=5.9 \pm 1.3^{+1.1}_{-0.7}~{\rm GeV^{-2}}$, 
for a mean value of the mass of the final hadronic system of $10~{\rm GeV}$ \cite{barberis}.
 Also results in proton dissociation events were presented at this conference. 
In $\rho^0$ photoproduction 
events with proton dissociation, the exponential slope is $b\simeq 5~{\rm GeV^{-2}}$ at $t \simeq 0$ \cite{sacchi}
and becomes flatter, $b \simeq 2.5 ~{\rm GeV^{-2}}$, at high $|t|$ \cite{piotr}.

In summary, vector meson production at HERA is a rich field for studying the interplay between soft
and hard interactions in diffractive processes. 
More luminosity and the forward proton spectrometers installed in both
the H1 and ZEUS experiments will allow to make more precise measurements.

\section*{Acknowledgments}
We wish to thank all the colleagues who participated in this parallel session,
our colleagues convenors of the shared sessions, the secretariat of the session, 
and the organizers and the secretariat of DIS96 for the warm hospitality.
VDD would like to thank PPARC and the Travel and Research Committee of the
University of Edinburgh for the support.
EG wants to thank G.~Barbagli, M.~Arneodo and A.~Levy for a careful reading of 
part of the manuscript.


\end{document}